\documentclass[aps,twocolumn,showpacs,showkeys,preprintnumbers,superscriptaddress,nobibnotes,floatfix,longbibliography,notitlepage,nofootinbib]{revtex4-1}

\pdfoutput=1
\usepackage{amsmath}
\usepackage{color}
\usepackage{graphicx}
\usepackage{xcolor}
\usepackage[colorlinks=true,allcolors=blue]{hyperref}
\usepackage{multirow}
\usepackage{float}
\begin{document}

\title{Wilks's Theorem,  Global Fits, and Neutrino Oscillations}
\author{J.M.~Hardin}
%\author{John Hardin}
\email{jmhardin@mit.edu}
%\email{zzzz}
\affiliation{Dept.~of Physics, Massachusetts Institute of Technology, Cambridge, MA 02139, USA}
%\affiliation{zzzz}

\date{\today}
\begin{abstract}

Tests of models for new physics appearing in neutrino experiments often involve global fits to a quantum mechanical effect called neutrino oscillations.
This paper introduces students to methods commonly used in these global fits starting from an understanding of more conventional fitting methods using log-likelihood and $\chi^2$ minimization.  Specifically, we discuss how the 
$\Delta\chi^2$, which compares the $\chi^2$ of the fit with the new physics to the $\chi^2$ of the Standard Model prediction, is often interpreted using Wilks's theorem.  This paper uses toy models to explore the properties of $\Delta\chi^2$ as a test statistic for oscillating functions.  
The statistics of such models are shown to deviate from Wilks's theorem.  Tests for new physics also often examine data subsets for ``tension'' called the ``parameter goodness of fit''.  In this paper, we explain this approach and use toy models to examine the validity of the probabilities from this test also.   Although we have chosen a specific scenario---neutrino oscillations---to illustrate important points, students should keep in mind that these points are widely applicable when fitting multiple data sets to complex functions.

\end{abstract}
\maketitle
%\end{CJK*}

\section{Introduction}

Models of new physics often make several related predictions that may not be testable in a single experiment.  Testing the model then requires integrating information from multiple types of experiments to perform convincing searches: so-called ``global fits''.    This paper introduces students to approaches commonly used in global fits to data in search of new physics, and provides cautionary examples of where implicit assumptions can break down.

An example of global fitting is the case of searches for ``sterile'' neutrinos---neutrinos that have no Standard Model interactions.   These are predicted to be partners of the three known active neutrinos, the $\nu_e$, $\nu_\mu$ and $\nu_\tau$.   If one sterile neutrino ($\nu_s$) is added to the theory, we call this model ``3+1.''  Such a model predicts that $\nu_e \rightarrow \nu_s$, $\nu_\mu \rightarrow \nu_s$ and $\nu_\mu \rightarrow \nu_e$ oscillations will all occur with the same mass splitting, which corresponds to a spatial frequency for monoenergetic neutrinos.  The spatial frequencies of these types of experiments are not expected to be independent in a 3+1 model.  Therefore, we must combine data from each type of oscillation experiment, and from multiple experiments of each type.    
The typical way to do this is to combine the model likelihood of the experiments and then maximize to find the most likely model.  Assuming the experiments are independent, this is as simple as adding together the log-likelihoods.

In this paper, we use the example 3+1 oscillations to explore statistical issues involved in comparing and combining experiments.  In particular, we look at
a commonly used approach, interpreting the $\Delta \chi^2$ using Wilks's theorem, typically applied in global fits to sterile neutrino models to determine limits and allowed regions.  We test the reliability of Wilks's theorem by throwing many fake experiments, which allows a ``frequentist test'' of the true probability.  ``Throwing'' data involves using a random number generator to create fake data with the same variations we expect in our experiments.  We also describe the ``PG-test,'' or ``Parameter Goodness of Fit test'', a method of measuring internal tension within the combined data sets that is commonly used in neutrino physics, and examine whether the $\chi^2$ probability accurately describes the true probability in the case of oscillations.

The goal of this paper is a pedagogical examination of commonly used statistical techniques and their reliability when fitting to models that involve oscillations. For this, we need to use example data sets.
Experimental data from oscillation experiments can have many complex features arising from systematic uncertainties.   For a review of the many issues involved with actual data, see Ref.~\cite{https://doi.org/10.48550/arxiv.2211.02610}.   To avoid distraction from these issues, and for educational purposes, it is simpler to use ``toy models'', where we control all aspects of the ``experiments.''   This allows the statistical effects to be clearly demonstrated.

\section{The Log-likelihood, $\Delta \chi^2$,  Wilks's Theorem, and Common Test Statistics}

In this section, we will give a brief overview of the log-likelihood, $\Delta \chi^2$,  and Wilks's theorem, focusing on what is relevant to the toy models presented in this paper.  It is not intended as a full review of statistics (A full statistical treatment is found in~\cite{bevington2003data}), but establishes the concepts needed for this discussion.

When looking at the results of an experiment, one vital piece of information is ``How likely is your data given your model''.  This is called the ``likelihood''.  A model of the data consists of a function telling us the probability of a given event, often one that has a set of parameters that can vary.  This function is known as a ``probability density function'', and the parameters are usually represented by a vector $\vec{\theta}$.  Assuming each data event is independent of the next, the likelihood can be found by finding the probability of each event and multiplying:

\begin{equation}
    L(\vec{x},\vec{\theta}) = \prod^n_i PDF(x_i,\vec{\theta}).
\end{equation}
where $\vec{x}$ is the set of data, $PDF$ is the probability density function representing the probability of a given event, $\vec{\theta}$ are the parameters of that function, and $L$ is the likelihood of the set of data.\footnote{Formally, the product of probabilities will be 0 for continuous-valued $PDF$s.  We won't discuss the reasons these infinitesimal elements can be ignored here.  For most physics applications, they do not affect the conclusion.  $L$ represents this product after dealing with this detail.}
$L$ is the number that contains the most information about the data's relation to the model, though its interpretation often requires approximations.  $L$ it is often unwieldy and computationally difficult to deal with, and so we almost always compute the likelihood in log-space, producing the log-likelihood:

\begin{equation}
    LL(\vec{x},\vec{\theta}) = \sum^n_i ln(PDF(x_i,\vec{\theta})).
\end{equation}

One especially relevant model is the normal distribution.  It is often possible to reduce the expected outcome of a given experiment to a series of normal distributions.  Indeed, any experiment that relies on counting a large number of events in a set of bins can be represented by a set of normal distributions\footnote{Formally, the expected distribution is a Poisson, but a Poisson distribution which expects $n$ counts is well approximated by a normal distribution with mean $n$ and standard deviation of $n^{\frac{1}{2}}$ when $n$ is large.}\footnote{Even when these bins are not completely independent, to the extent that the dependence is described by a correlation matrix, a set of normal distributions still applies.  We won't be talking about correlated data in this paper, however}:

\begin{eqnarray}
    LL(\vec{y},\vec{\theta}) &&= \nonumber  \\ 
    \nonumber &&= \sum^d_j ln\left(\frac{1}{\sigma_j\sqrt{2\pi}}e^{\frac{-(y_j-\mu_j)^2}{2\sigma_j^2}}\right) \\
    \nonumber &&= \sum^d_j \frac{-(y_j-\mu_j)^2}{2\sigma_j^2} + ln\left(\frac{1}{\sigma_j\sqrt{2\pi}}\right), \label{eq:gauslikelihood}
\end{eqnarray}
where $\vec{y}$ are the data, $\vec{\mu}$ are the mean expectations of the model (dependent on $\vec{\theta}$), and $\vec{\sigma}$ are the standard deviations of each element.  One may recognize a portion of the first term as the traditional $\chi^2$.  In principle, $\vec{\sigma}$ can depend on $\vec{\theta}$ as well, but it often has minimal dependence\footnote{When looking for new physics, the effect is often relatively small, and the effect on the standard deviation is even smaller.}.  Since we are most concerned with differences in $LL$, if:
\begin{equation}
    \chi^2 = \sum^d_j \frac{-(y_j-\mu_j)^2}{\sigma_j^2}~,
    \label{eq:chi2def}
\end{equation}
then
\begin{equation}
    LL(\vec{y},\vec{\theta}) \approx \frac{-\chi^2(\vec{y},\vec{\theta})}{2}.
\end{equation}

\begin{figure}[tb]
    \centering
    \includegraphics[width=\linewidth]{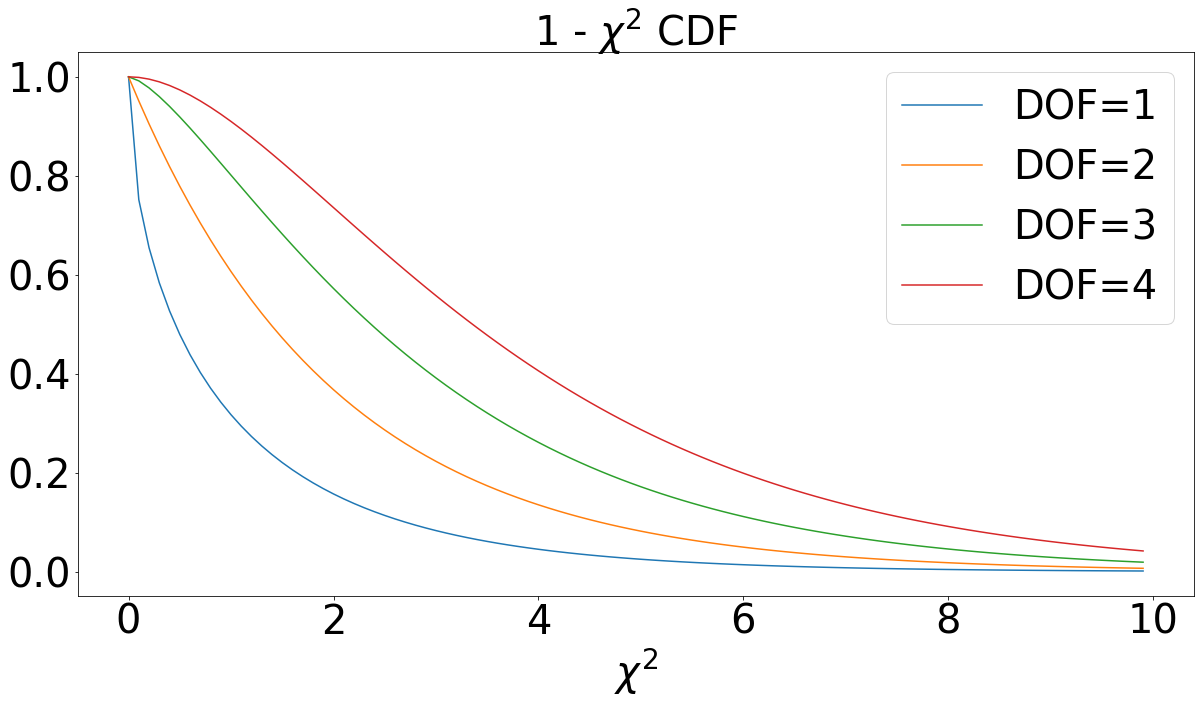}
    \caption{1 minus the CDF of a $\chi^2$ distribution with various degrees of freedom. \label{fig:chi2cdf}}
\end{figure}

The $\chi^2$ as defined by Eq~\ref{eq:chi2def} has a major advantage as a test statistic.  
If the data do in fact follow the underlying normal distribution, then the cumulative distribution function (CDF) of that calculated $\chi^2$ is easy to compute.  We index these distributions based on the number of underlying normal distributions ($d$ in Eq~\ref{eq:chi2def}).  
When a variable is $\chi^2$ distributed, we say it follows a $\chi^2$ distribution with $d$ degrees of freedom.  Larger values of $\chi^2$ are progressively less likely, and the ease of calculating the CDF means we can know the probability of seeing a $\chi^2$ that large or larger.  
We call this the``$p$-value'' associated with that $\chi^2$.  Different fields consider different levels of $p$-value significant, but in physics, we typically think that things are ``interesting'' around $p < \frac{1}{300}$ and consider ourselves to have definitively seen something around $p < \frac{1}{2 \times 10^6}$
\footnote{These values correspond to a deviation from the mean of $3\sigma$ and $5\sigma$ for a 1 dimensional normal distribution distribution.}  The distribution is plotted for the first few degrees of freedom in Fig~\ref{fig:chi2cdf}.

Analyses, however, often aim to choose the ``best'' fitting model among the range of represented models.  To do this, one finds the maximum loglikelihood:

\begin{equation}
    max_{\vec{\theta}} ~ LL(\vec{y},\vec{\theta})~,
\end{equation}
or, equivalently, the minimum $\chi^2$:
\begin{equation}
    min_{\vec{\theta}} ~ \chi^2(\vec{y},\vec{\theta}).
\end{equation}

In addition to asking ``what is the best model'', we usually want to ask ``how much better is it than the null model'' (the model with no new physics).  If we label our parameters corresponding to the null model as $\vec{\theta}_0$, then we can define:

\begin{equation}
    \Delta LL(\vec{y}) = max_{\vec{\theta}} ~LL(\vec{y},\vec{\theta}) - LL(\vec{y},\vec{\theta}_0)~,
\end{equation}
and
\begin{equation}
    \Delta \chi^2(\vec{y}) = \chi^2(\vec{y},\vec{\theta_0}) -  min_{\vec{\theta}} ~\chi^2(\vec{y},\vec{\theta}),
    \label{eq:delchi2def}
\end{equation}
Then, to excellent approximation,
\begin{equation}
    \Delta LL(\vec{y}) = \frac{\Delta \chi^2(\vec{y})}{2}.
    \label{eq:delchi2}
\end{equation}

Equation~\ref{eq:delchi2} is very important\footnote{We have derived Equation~\ref{eq:delchi2} by making a series of approximations to Normal Distributions and using fixed errors.  This was done primarily for ease of understanding, but Wilks's theorem is more general and the equation applies to models more complex than a sum of Normal Distributions.} Under a certain set of assumptions, it is a test statistic that will follow a $\chi^2$ distribution with $DOF=dim({\vec{\theta}})$, the number of model parameters~\cite{Wilks:1938dza}.  This fact is known as Wilks's theorem and is used widely as a simplifying assumption in physics. 
From the $\chi^2$ distributions shown in Fig.~\ref{fig:chi2cdf} one can see that adding additional parameters will decrease the $p$-value for a given fit.  Intuitively, this is accounting for the better fit produced merely by increased model flexibility.  
The assumptions of Wilks's theorem are very formally stated, but, broadly, they require that: the parameters of the model affect the prediction in independent ways, the parameters have a full range over which to vary, and that there are many independent data points relating to the same model.
For many models within physics this is true, or approximately true across most of parameter space.
When this applies, it is, computationally, very  convenient, as it requires that the analyzer only fit the original set of data

Unfortunately, not all models satisfy the assumptions of Wilks's theorem.  Many models, especially physical models, can have limits on the allowable parameter space (such as an amplitude being positive-definite).  In the case of oscillating functions, the frequency parameter of the function can scan over multiple independent local minima in $\Delta\chi^2$ and result in model flexibility that is not accounted for by one additional parameter.  We are examining this effect on this physically motivated model, but an in-depth treatment of various violations of the assumptions of Wilks's theorem can be found in Ref.~\cite{Algeri2020}.
The models will behave as if one were checking many independent models and taking the best, which will produce a larger $\Delta \chi^2$ and therefore a smaller $p$-value than warranted.  
To understand this CDF of the $\Delta \chi^2$ of this model, it is necessary to perform statistical trials, which involves throwing fake data in accordance with your null-model.  The $\Delta\chi^2$ is then calculated for each of these sets of fake data, and the distribution of that test statistic can be used to assess the significance of the result.  

Both a model which does and a model which does not follow Wilks's theorem are presented below.  A method for understanding the significance of a given test statistic independent of the theorem is presented as well.

\section{Defining the Toy Models}

One toy model we will examine is relatively well described by Wilks's theorem, while the other is not.  We will structure the parameters of the first in a slightly odd way to maintain the analogy to the second.  To understand this, we start with the parameters of the oscillatory model, which is the poorly described one.

\subsection{The Oscillatory Model\label{sec:oscmodel}}

In this paper, for our oscillatory model, we use the example of neutrino oscillations.  Very briefly, neutrino oscillations refer to the tendency of neutrinos produced in one flavor to have some probability to either be detected in another flavor (``appearance'') or to not be detected at all (``disappearance'') after some time has elapsed.   Since detectable neutrinos are produced with some energy $E$, over time $t$ they must travel some distance $L$.  If neutrinos' flavor states are not the same as their mass states, then the change in probability for flavor detection is periodic in 
\begin{equation}
x=L/E, 
\end{equation}
This effect has been observed among the three known neutrino flavors, and the Nobel Prize was awarded for neutrino oscillation results in 2015~\cite{nobelpress2015}.  The recent decades have seen a large diversity of neutrino experiments.  For a full review of the current neutrino theory and experimental methods, please see Ref.~\cite{Athar_2022}.

This periodicity is also the smoking gun for new types of oscillations, including 3+1 models which look for a proposed new type of neutrino rthat only interacts with the Standard Model through oscillations.  However, this periodicity is also the cause of statistical problems when neutrino data are analyzed too simplistically.  In this paper, we will use toy models to explore the effects of interest, but students that would like to learn more about sterile neutrinos and oscillations can see Ref.~\cite{https://doi.org/10.48550/arxiv.1609.07803}.

The toy models presented in this article are set to mimic the behavior of the fits to 3+1 sterile neutrinos in~\cite{Diaz:2019fwt}.  That is, there are 3 amplitudes, corresponding to $e\rightarrow e$ disappearance, $\mu\rightarrow \mu$ disappearance, and $\mu\rightarrow e$ appearance.  These 3 amplitudes only have 2 free parameters between them.  In papers on global fits, you will see them called $|U_{e4}|^2$ and $|U_{\mu 4}|^2$, but to make this paper more readable, we will abbreviate these as $A_e$ and $A_\mu$.  Also in 3+1 models, there is a parameter that regulates the frequency which is called the mass splitting, usually written as proportional to the parameter $\Delta m^2$; but here, for simplicity we will use $m$ for that term.  The bottom line for this discussion is that the model we will want to fit looks like this:
\begin{equation}
    \nu_e \rightarrow \nu_e : 1-4(1-A_e)A_e \sin^{2}(mx)
    \label{eq:eemodel}
\end{equation}
\begin{equation}
    \nu_\mu \rightarrow \nu_\mu : 1-4(1-A_\mu)A_\mu \sin^{2}(mx)
    \label{eq:uumodel}
\end{equation}
\begin{equation}
    \nu_\mu \rightarrow \nu_e : 4 A_e A_\mu \sin^2(mx).
    \label{eq:uemodel}
\end{equation}

Other than a scale factor, $x$ behaves directly as $\frac{L}{E}$ in a physics 3+1 sterile neutrino model.  These models further have the property that, due to $0\le A\le 1$, the amplitudes are strictly positive.  This will lead to a leftward bias of the test statistic PDF,
as opposed to the rightward bias exhibited 
by checking many independent frequencies and taking the best.

In order to represent the 3 classes of experiment, we use a toy model consisting of 2 sets of ``disappearance'' data, reflecting Eqs.~\ref{eq:eemodel} and \ref{eq:uumodel}, and 1 set of ``appearance'' data, reflecting Eq.~\ref{eq:uemodel}.  The disappearance data is modeled as a set of ratios in $n$ independent bins, each with error described by a normal distribution, whose uncertainty is the same.  The appearance data is modelled with a large Poisson distribution.  These are easy to generate, and relatively easy to fit using standard optimization tools (simplex, {\it etc}).   However, in the oscillatory model such as this one, there are many false minima, and one must be careful to fit from a variety of starting points.

\subsection{The Linear Model\label{sec:linearmodel}}

We want to compare and contrast the oscillatory model results to a model that conforms to Wilks's theorem.
An example of such a model is one where we will fit purely a slope to the data.  
Let us define the following function:
\begin{align*}
    f(A) &= &\nonumber\\
    &\frac{5}{2}A & A < \frac{2}{5}, \\
    &1 - \frac{10}{3}(A - \frac{2}{5}) & A > \frac{2}{5}.\label{eq:appf}
\end{align*}

$f$ is defined this was so that the models can vary from positive to negative in amplitude.  Our A's range from 0 to 1, but $f(A)$ varies from -1 to 1 while retaining the property $f(0) = 0$.

Then we can invent a new model (not a model that is used in neutrino physics, just a simple example) that we define as follows:
\begin{equation}
    \nu_e \rightarrow \nu_e : 1-f(A_e )(x-x_{mid}) \label{eq:eelin}
\end{equation}
\begin{equation}
    \nu_\mu \rightarrow \nu_\mu : 1-f(A_\mu )(x-x_{mid})   \label{eq:uulin}
\end{equation}
\begin{equation}
    \nu_\mu \rightarrow \nu_e : 4f(A_e)f(A_\mu)(x-x_{mid}),   \label{eq:uelin}
\end{equation}
where $x_{mid}$ is the midpoint of the x range.
To make the comparison clear, we keep the amplitude parameters connected as in the oscillatory case, but now we have removed any frequency dependence (there is no $m$ in this model). 

\begin{figure}[tb]
    \centering
    \includegraphics[width=\linewidth]{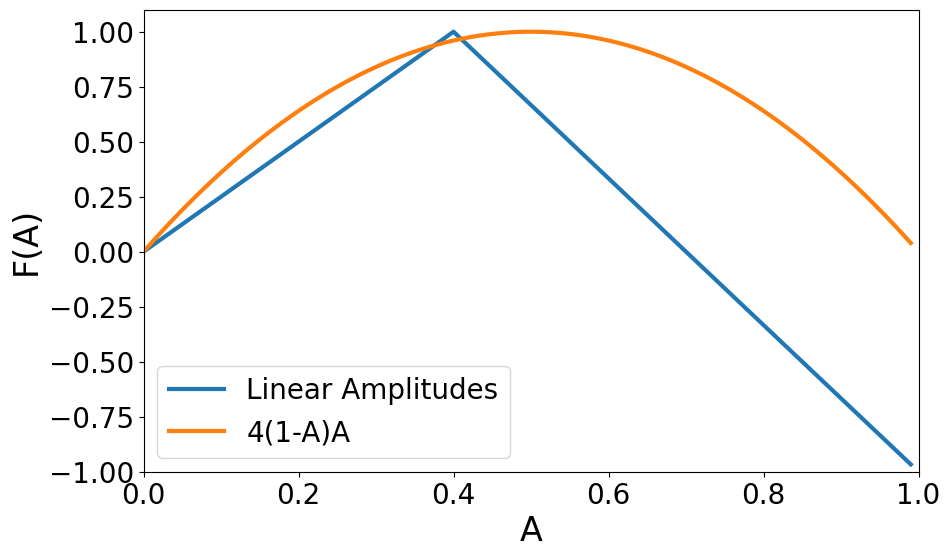}
    \caption{A plot comparing the response $f(A)$ for the ``disappearance'' case of the linear model above (Eq.~\ref{eq:eelin}) and the ``disappearance'' case of the sterile neutrino model (Eq.~\ref{eq:eemodel}) where ``$f(A)$''$=4(1-A)A$.  \label{fig:ModelA}}
\end{figure}

This function is compared to the equivalent from the oscillatory model in Figure~\ref{fig:ModelA}.  Note that the oscillatory function is strictly positive - this bound is one source of the deviation from Wilks's theorem as is discussed later.

\section{An Additional Statistic: PG-tension\label{sec:pgdescribe}}

We will also be evaluating both models using the ``parameter goodness of fit'' (PG) test as defined in Ref.~\cite{PhysRevD.68.033020}.   This is commonly used within neutrino physics, especially in global fits to sterile neutrinos, although is less well-known outside of the neutrino community.   It is a very useful test statistic that can be applied to any type of global fit.
The PG test is a way to evaluate the robustness of a model.  The PG test involves splitting the data into multiple sets with independent models, fitting these models independently, and then comparing how much the smaller sum of the individual $\chi^2$'s is than the combined $\chi^2$ as in Eq.~\ref{eq:chipg}.

For the case of sterile neutrino oscillations,  the division is usually along the lines of ``disappearance experiments,'' which should fit to Eqs.~\ref{eq:eemodel} and \ref{eq:uumodel}, and ``appearance experiments'' which should fit to Eq.~\ref{eq:uemodel}.    One can fit the two sets of data independently.  The disappearance data sets have three free parameters in $A_e^{dis}$, $A_\mu^{dis}$, and $m_{dis}$.    The appearance data has only two free parameters $(A_e A_\mu)^{app}$, and $m_{app}$.   The results of these two fits are compared to the global fit, which has three free parameters $A_e^{glob}$, $A_\mu^{glob}$, and $m_{glob}$.   If the data sets agree internally then the results of the fits will agree.  That is, if:
\begin{eqnarray}
&A_e^{dis}=A_e^{glob},  \notag \\ 
&A_\mu^{dis}=A_\mu^{glob}, \notag \\
&(A_e A_\mu)^{app}=( A_e^{dis} A_\mu^{dis}) =(A_e^{glob} A_\mu^{glob}),  \notag \\
&{\rm and}~m_{dis}=m_{app}=m_{glob},
\end{eqnarray}
then there is perfect agreement.

The level to which these parameters do not agree is called ``PG tension.''   One finds the tension in the case of 3+1 in the following way:
\begin{equation}
    \chi^2_{PG} = \chi^2_{glob} - \chi^2_{app} - \chi^2_{dis},  \label{eq:chipg}
\end{equation}
where the $\chi^2$ values are minimum from the individual fits.

We have described this test statistic based on its most common use in global fits,  but it is applicable to any datasets that should share an underlying model.    Thus, this is an extremely useful tool for understanding internal agreement of data.   Therefore, we will explore this test statistic further.
Under somewhat similar assumptions to Wilks's theorem, we expect $\chi^2_{PG}$ to follow a $\chi^2$ distribution with $N_{PG}$ degrees of freedom:
\begin{equation}
    N_{PG} = N_{app} + N_{dis} - N_{glob}  =  2 + 3 - 3,   \label{eq:npg}
\end{equation}
where the $N$'s are the number of parameters in each fit.
A PG test with a small $p$-value suggests that the model does not describe the data in an internally consistent fashion.  Choosing which data to put in which division is necessarily somewhat arbitrary and the selection may not conform to relevant degrees of freedom.  In the sterile neutrino case, it is natural to separate appearance and disappearance as they are expected to be the most independent.  A very serious problem in neutrino physics related to the 3+1 models is that the PG test $p$-values are consistently very low \cite{https://doi.org/10.48550/arxiv.2211.02610}, which is leading to controversy in the field.  However, in this case, as we are using toy-models, we know that the data are internally consistent, so we can evaluate the applicability of this test.

Similar to the normal fit case, we might expect these assumptions to be violated by the oscillatory model and to be upheld by the linear case.  We can test this conclusion with trials in the same way.  However, the picture is more complicated in this case, as there are competing effects, which we will discuss in the results.   Thus, this is a very interesting case study.

\section{Results of Fitting the Toy Models}

And now we will fit the models as defined to a large number of different datasets, randomly generated as defined.  In order to analyze the agreement with Wilks's theorem, we will histogram both the results (shown in red points) and the CDF of the results (shown in blue points).  We will compare them to the lines of an ideal $\chi^2$ distribution with the appropriate numbers of degrees of freedom.  We will define ``conservative'' in the following way:  when the models tend to the left of the ideal distribution, Wilks's theorem is ``overly-conservative,'' and when they tend to the right, it Wilks's theorem is ``under-conservative.''    An over-conservative case will over-estimate the $p$-value, and the under-conservative case will under-estimate the $p$-value.  We name them this way because an overestimated $p$-value is less likely than warranted to reject the null (no new physics) model, while an underestimated $p$-value is more likely than warranted.  Full agreement appears when the number of fits at a given $\Delta \chi^2$ agree with the $\chi^2$ distribution.

\subsection{Simulation Specifications}

Obtaining these distributions requires defining the simulation parameters.  Each simulation assumes 3 different experiments, each with a number of bins $N_b$.  The x values of these bins run from 0.2 to 2.0 - these numbers are in rough agreement with common L/E bounds in sterile neutrino experiments.  

The two sets of bins corresponding to the electron and muon disappearance experiments are simulated identically.  As we are only simulating the null distribution, the disappearance experiments are simulated by throwing a random Normal Distribution with a mean of 1.0 and a standard deviation of 0.1.  Disappearance experiments typically have fairly large counts in each bin, and so the natural Poisson nature of counting is well approximated by a normal distribution.

The Appearance experiment is done differently.  The bins are assigned a background value of 1000 per unit, and the appearance is then amplified by a factor of 50 compared to Eqs.~\ref{eq:uemodel},\ref{eq:uelin}.  The amplification represents the detector acceptance that is typically divided out in the ideal case of the disappearance experiments, while the background simulates a reasonable experiment.  The Poissonian likelihood is used when calculating these likelihoods, but, as discussed above, it is still expected to conform to Wilks's theorem, and the bins have enough counts that the Poissonian is close to normal.

\begin{figure}[tb]
    \centering
    \includegraphics[width=\linewidth]{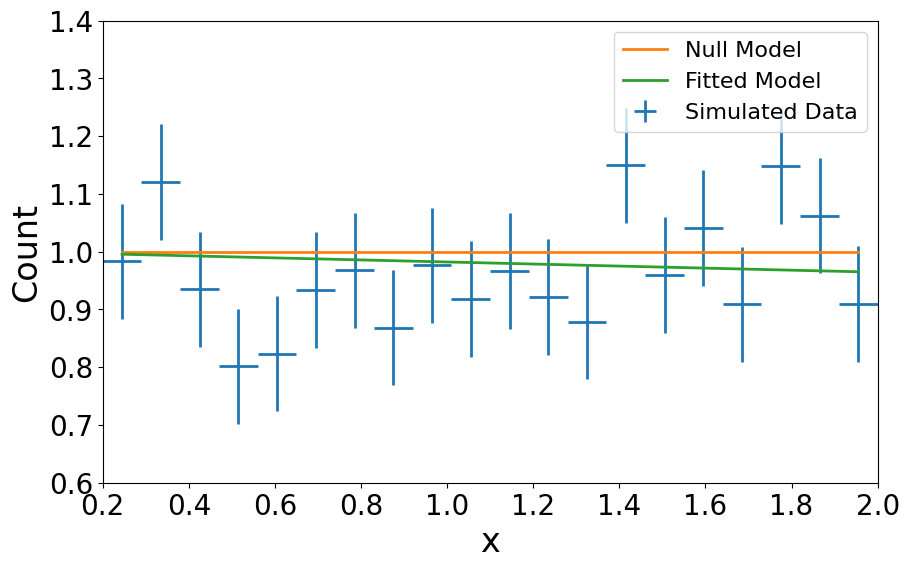}
    \caption{One single fit to 20 bins using the linear model.  The ``null'' is the fixed background, while the fitted model is the best fit linear model. \label{fig:linearsample}}
\end{figure}

For the linear model, only 20 bins were simulated for each experiment.  For the oscillatory model, we will examine both the 20 and 100 bin case.  Figures~\ref{fig:linearsample} and~\ref{fig:oscsample} show what one fit looks like for both types of models.

\subsection{Linear Model Results}

\begin{figure}[tb]
    \centering
    \includegraphics[width=\linewidth]{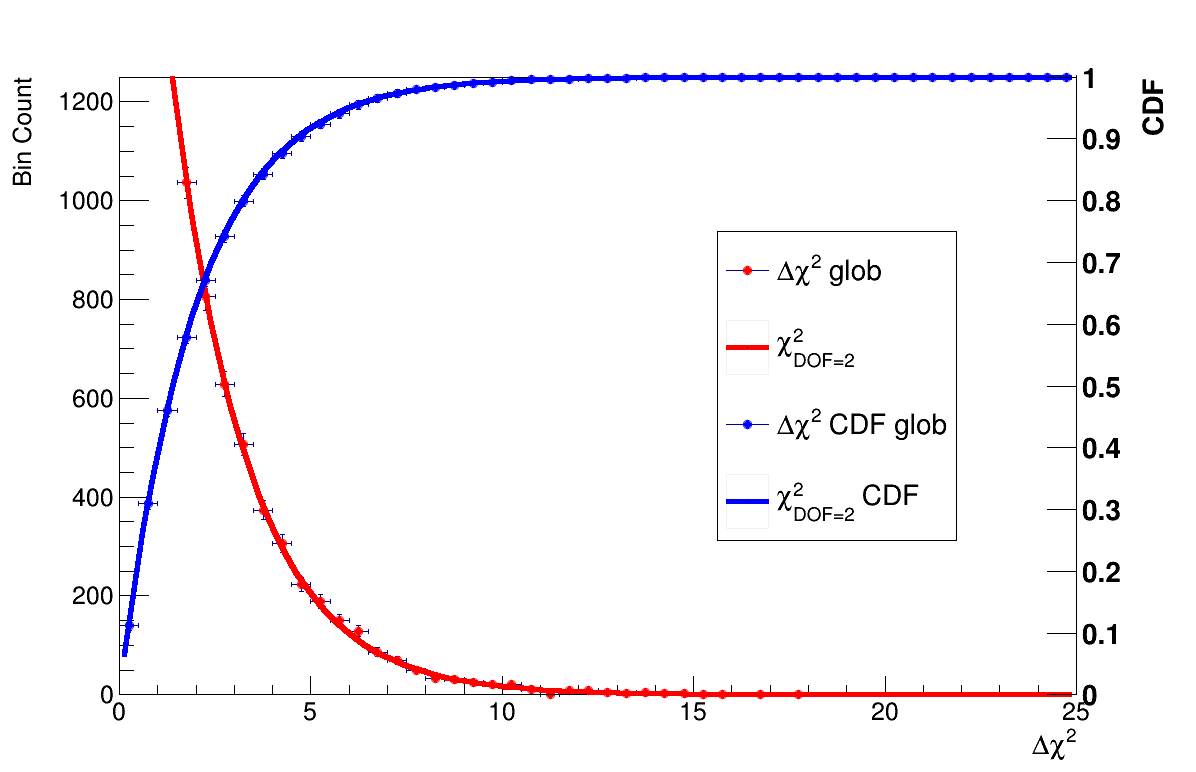}
    \caption{``global'' fits for the linear model.  The data can be seen to match the ideal lines.  Please note that while the vertical error bars correspond to the error on the bin, the horizontal error bars represent the bin width, as often seen in particle physics plots. \label{fig:linearglob}}
\end{figure}

We will start with the linear model as defined in Sec.~\ref{sec:linearmodel}.  Looking at the overall fit in Fig.~\ref{fig:linearglob}, we see excellent agreement with the expectation of 2 degrees of freedom.  The points from the trials line up within their error on the ideal curve (red points vs curve), as is also true for the  CDF of the results (blue points and curve).  Thus, in this case, where we expected Wilks's theorem to work well, it clearly does.

\begin{figure}[tb]
    \includegraphics[width=\linewidth]{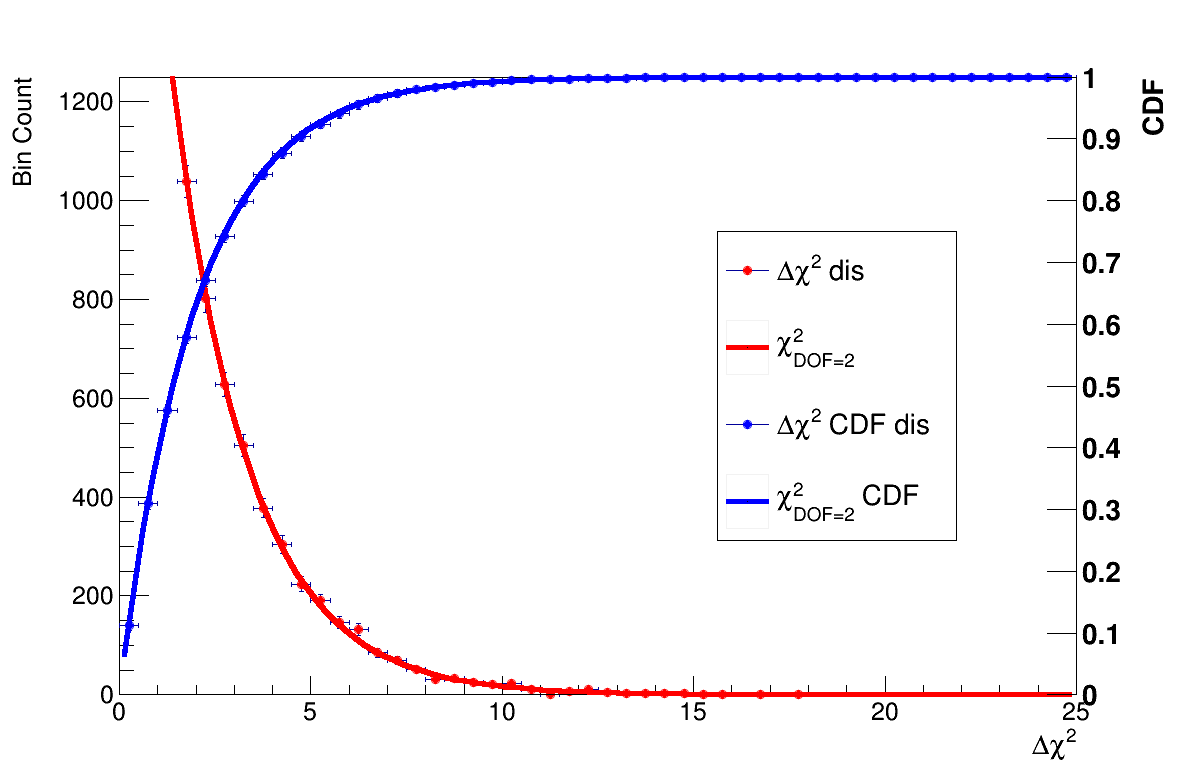}
    \includegraphics[width=\linewidth]{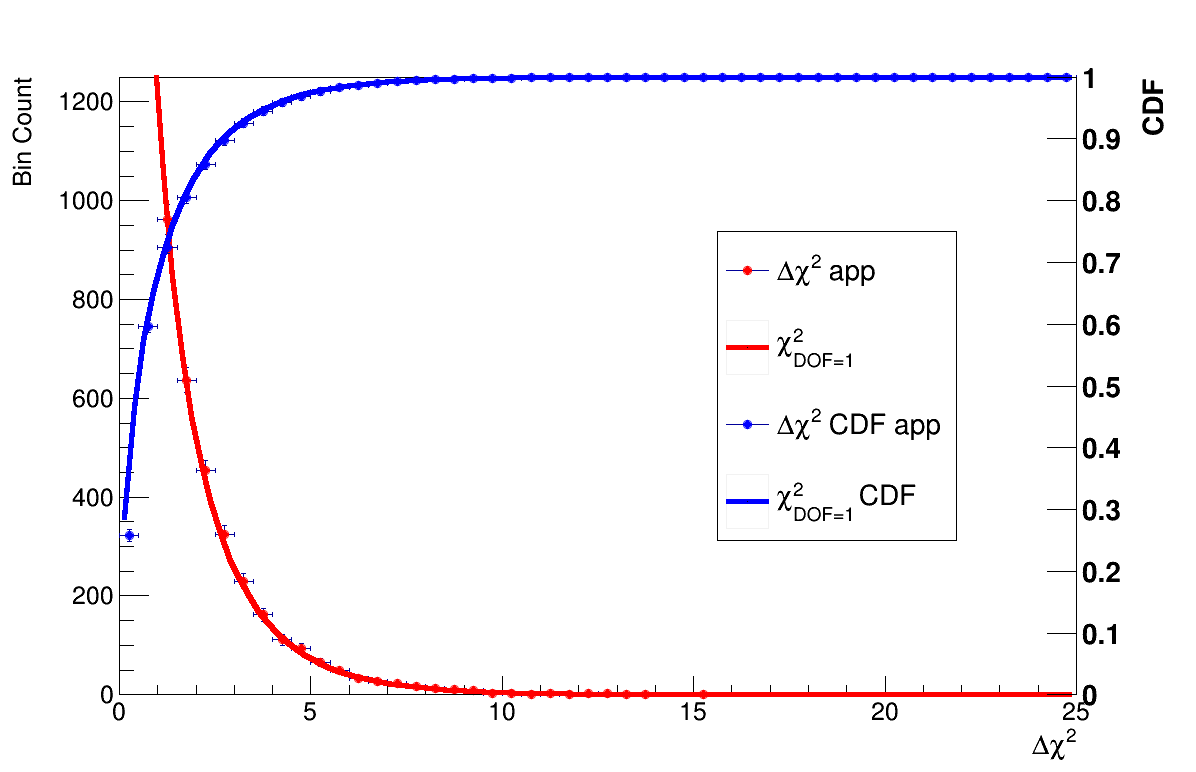}
    \caption{``disappearance'' (top) and ``appearance'' (bottom) fits for the linear model.  The data agree with the expectation. \label{fig:linearappdis}}
\end{figure}

In analogy with the oscillatory model discussed above, we break the linear model into ``disappearance'' (Eqs.~\ref{eq:eelin} and \ref{eq:uulin}) and ``appearance'' (Eq.~\ref{eq:uelin}) fits 
shown in Fig.~\ref{fig:linearappdis}.  
They show similar good agreement.  

Finally, we perform the PG test as described in Sec.~\ref{sec:pgdescribe} on the linear model.  
In this case, we need to count the number of parameters in the three fits (see Eq.~\ref{eq:npg}).
Note that the ``global'' and ``disappearance'' fit only 2 parameters in this case (the two amplitudes), while the ``appearance'' has 1 parameter.  Therefore, according to Eq.~\ref{eq:npg},  $N_{PG}=1$.  

\begin{figure}[tb]
    \includegraphics[width=\linewidth]{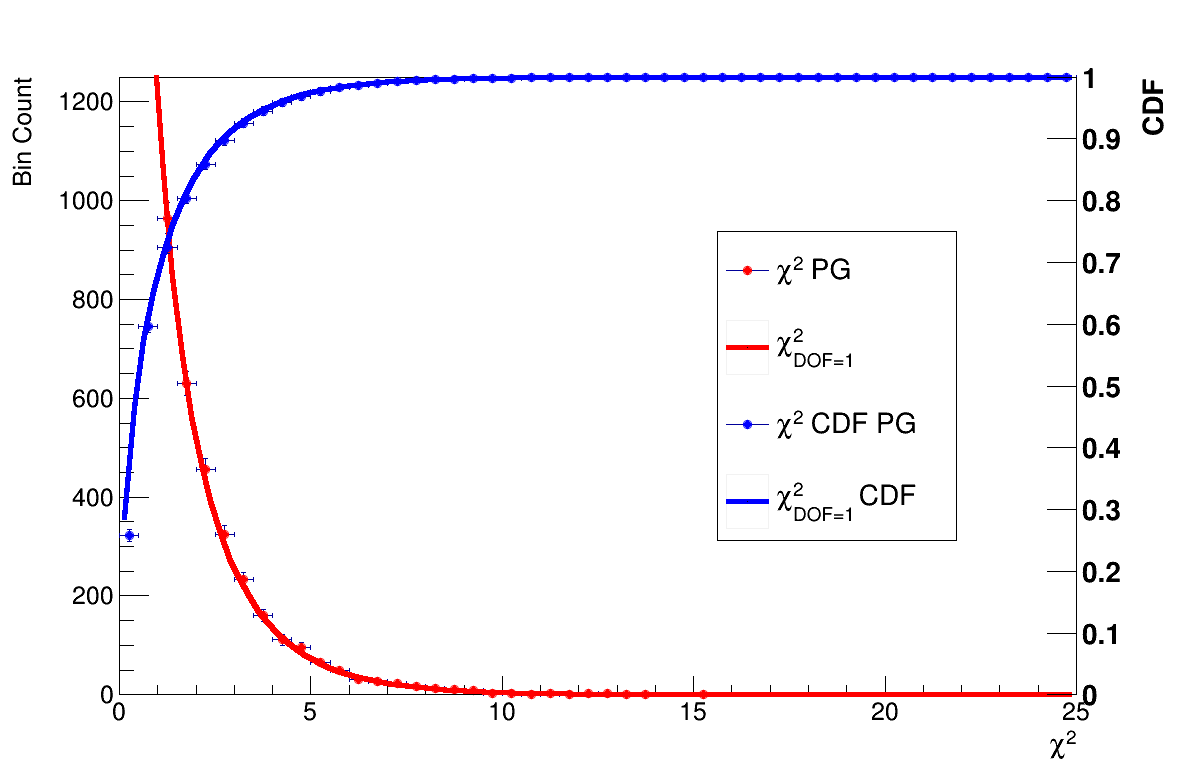}
    \caption{``PG'' fits for the linear model.  Once again, the linear model is in agreement with the ideal case.  \label{fig:linearpg}}
\end{figure}

We evaluate Eq.~\ref{eq:chipg}, to find the results shown in Fig.~\ref{fig:linearpg}.  The PG test is again in excellent agreement with the expectation.

From the figures shown, we can look see that the linear fit follows Wilks's theorem quite precisely, and the PG test follows the expected $\chi^2$.  Linear (and, more generally, polynomial) models often satisfy the assumptions of statistical theorems, and therefore their test statistics are usually well described by the theorems.

\subsection{Oscillatory Model Results}

\begin{figure}[tb]
    \centering
    \includegraphics[width=\linewidth]{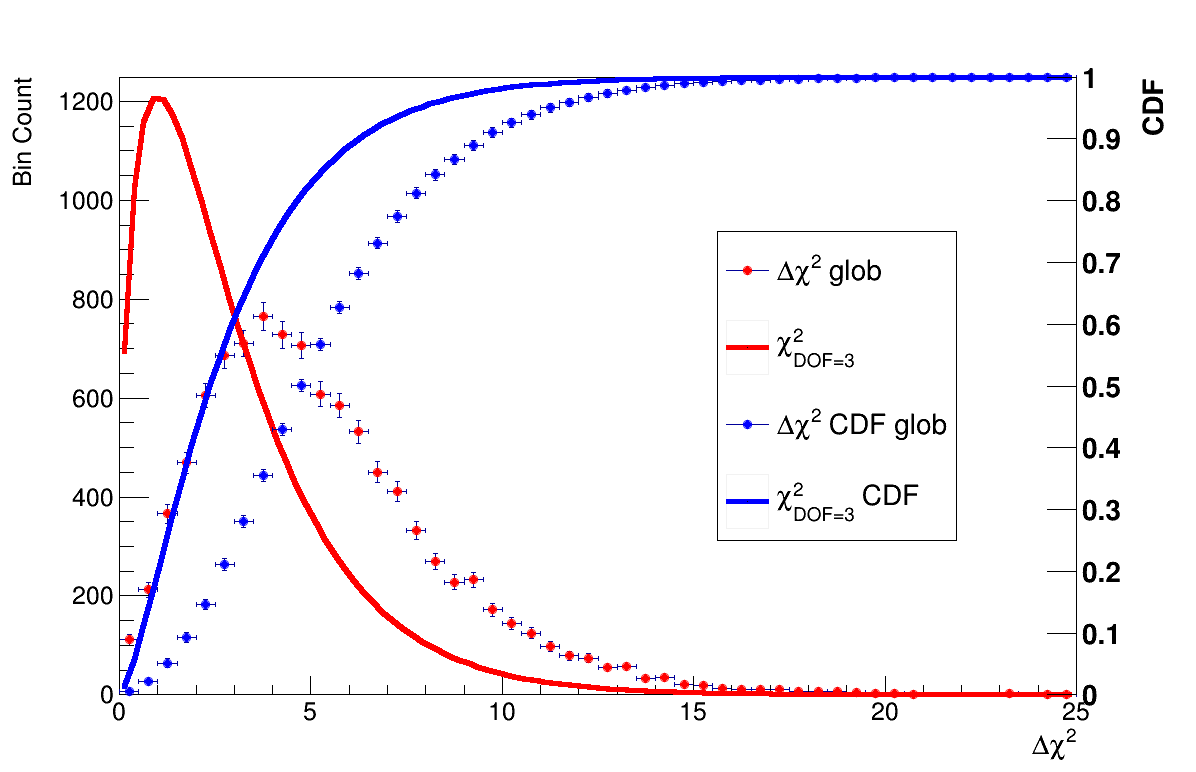}
    \includegraphics[width=\linewidth]{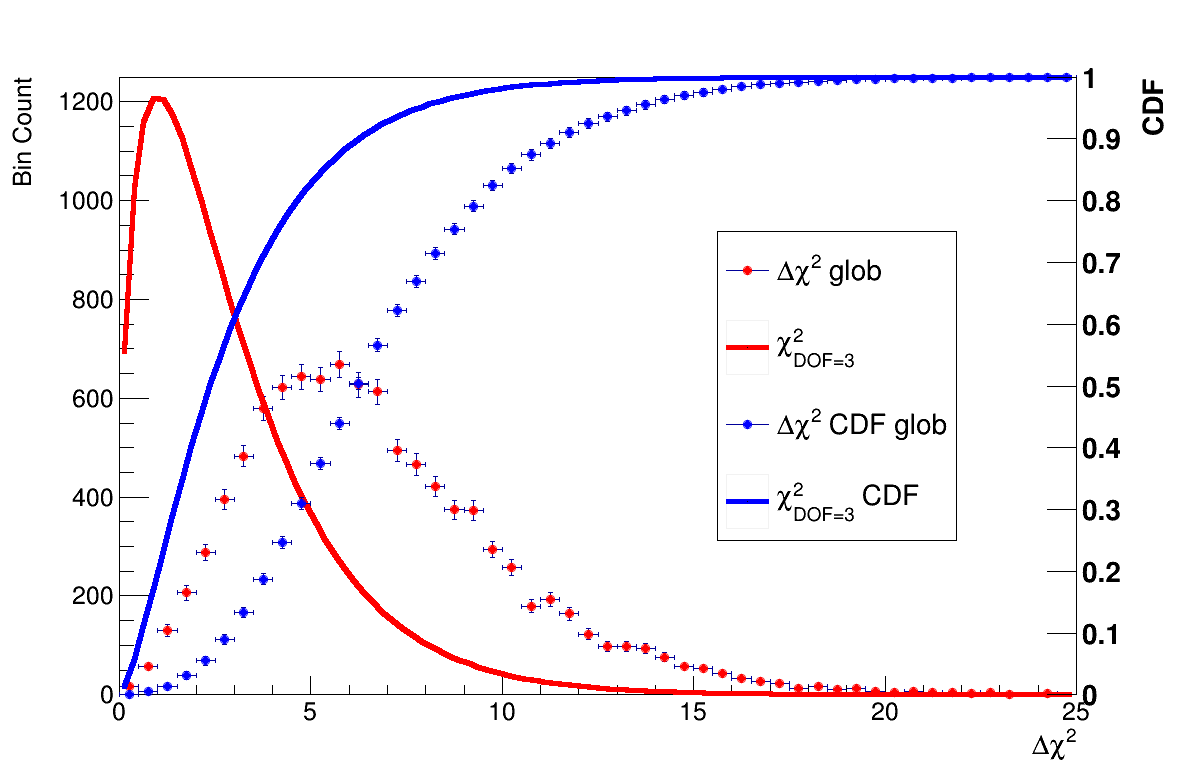}
    \caption{The ``global'' fit to the oscillatory model.  The $\Delta \chi^2$ is clearly to the right of the 3 degree of freedom line.  The top plot uses 20 data points in each of the 3 pseudo-experiments, while the bottom plot uses 100. \label{fig:delglobal}}
\end{figure}

In contrast to the linear model, the oscillatory model does not give results that even closely approximate a $\chi^2$ distribution.  To investigate the effect of experimental conditions on this fact, we ran trials with both 20 and 100 bins in each of the 3 ``experiments''.  Figure~\ref{fig:delglobal} shows the results for the overall fit, and it is clear that Wilks's theorem is under-conservative.  The probability mass is well to the right of the Wilks predicted line.

\begin{figure}[tb]
    \centering
    \includegraphics[width=\linewidth]{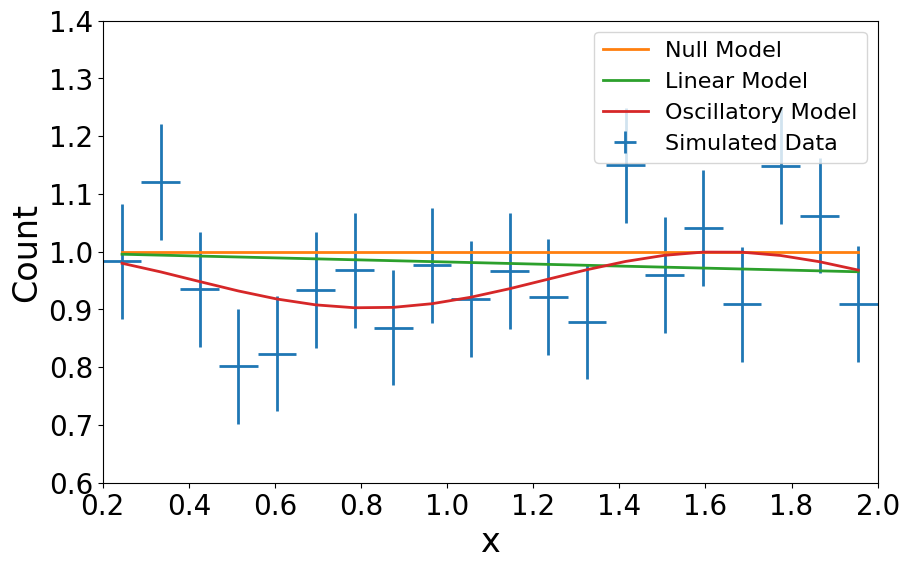}
    \caption{One single fit to 20 bins using the linear model and the oscillatory model.  The ``null'' is the fixed background.  These data are the same as Fig.~\ref{fig:linearsample}. \label{fig:oscsample}}
\end{figure}

\subsubsection{Appearance and Disappearancee}

\begin{figure}
    \centering
    \includegraphics[width=\linewidth]{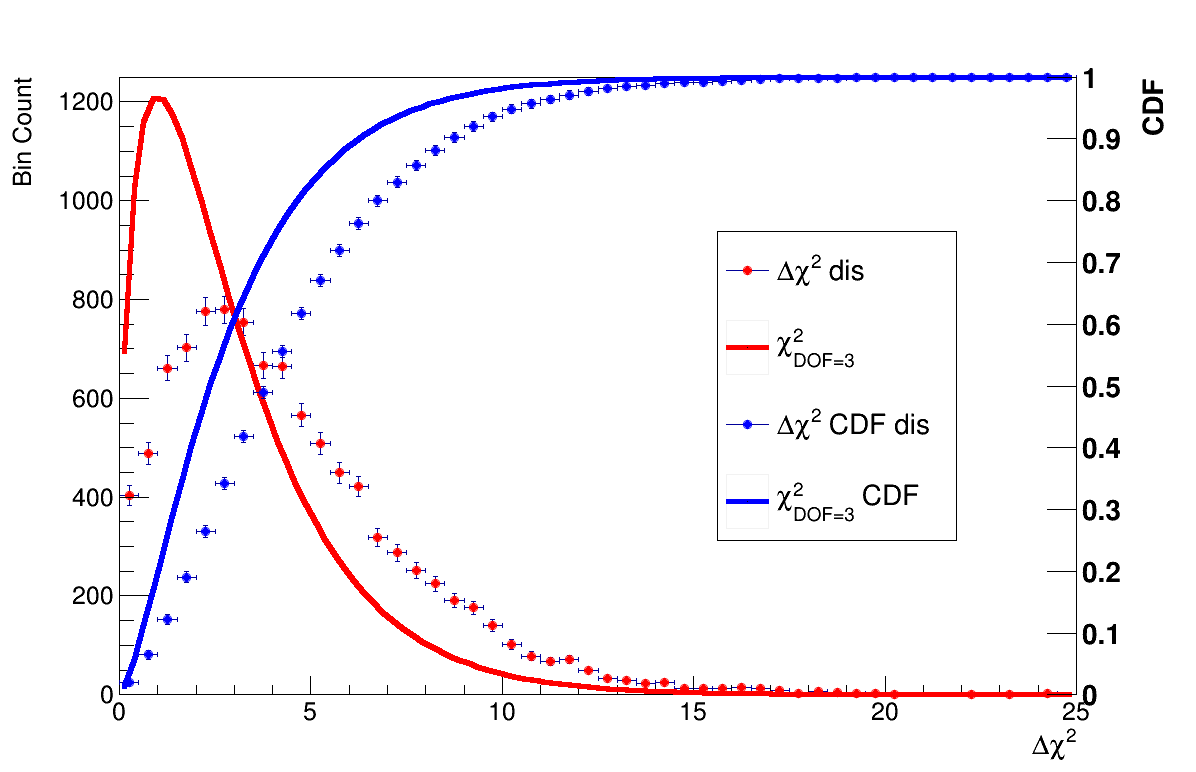}
    \includegraphics[width=\linewidth]{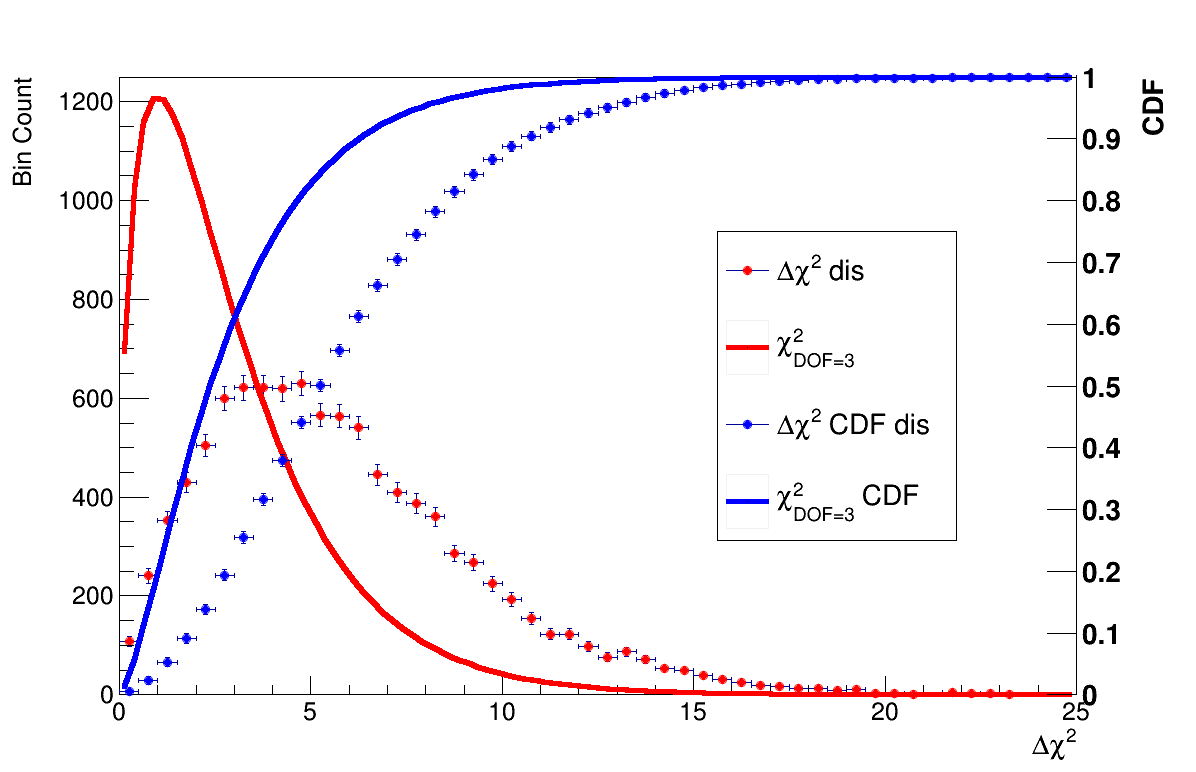}
    \caption{The ``disappearance'' portion of the toy fit.  This is fit without the appearance portion, but the mass squared difference is the same for the 2 kinds of disappearance.  The trials based statistics do not agree with a $\chi^2$ distribution. \label{fig:disglobal}}
\end{figure}

\begin{figure}
    \centering
    \includegraphics[width=\linewidth]{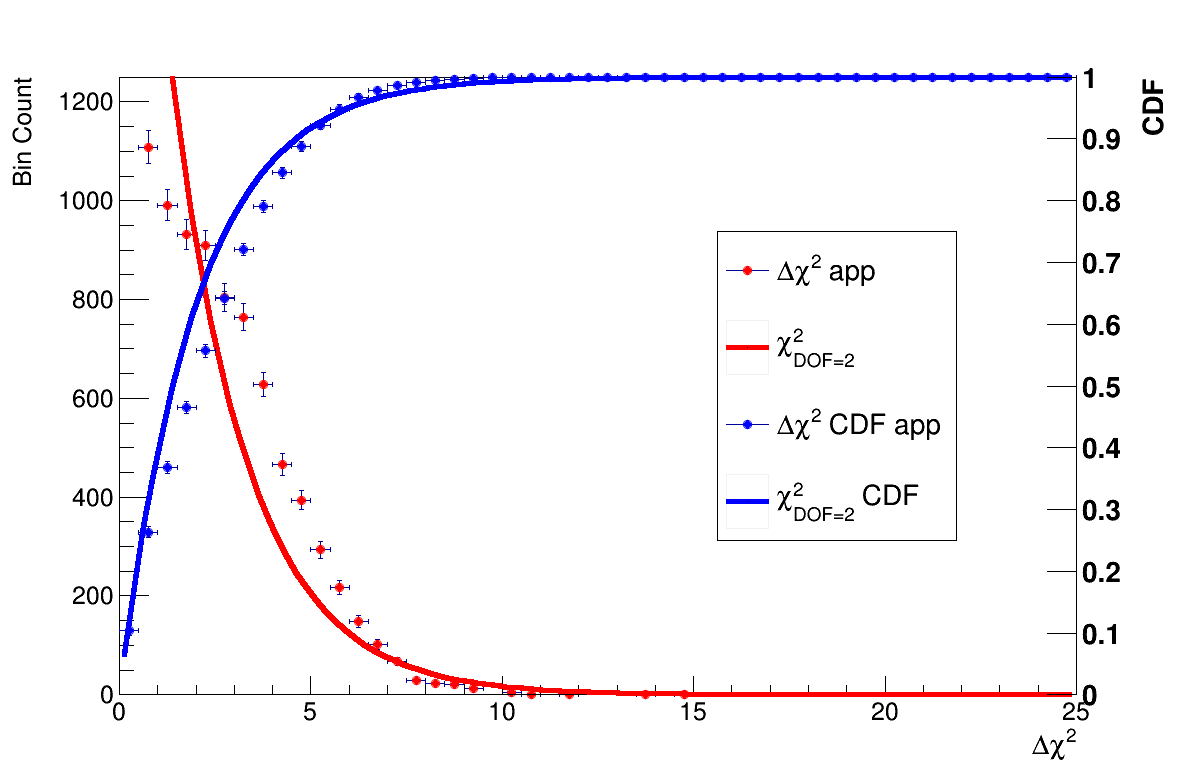}
    \includegraphics[width=\linewidth]{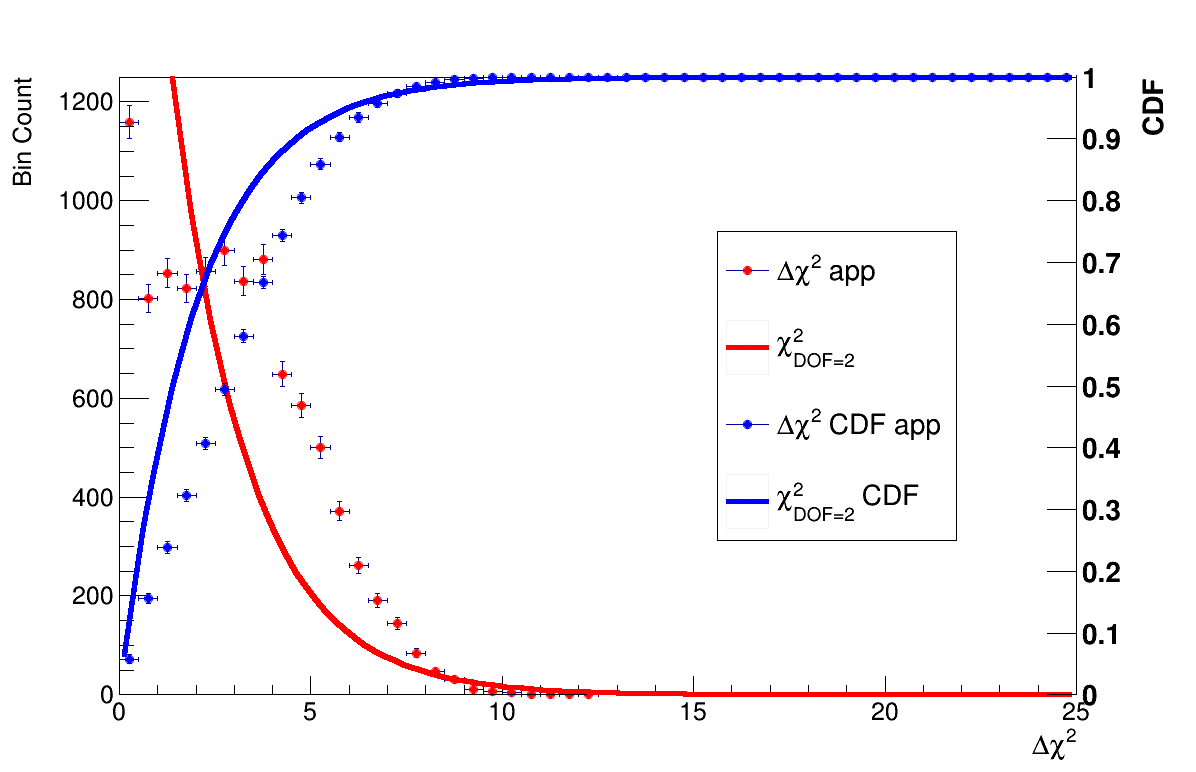}
    \caption{The ``appearance'' portion of the toy fit.  This is fit without the disappearance portions, and so only has 2 degrees of freedom.  Like the other oscillatory cases, Wilks's theorem does not give a good prediction of the test statistic. \label{fig:appglobal}}
\end{figure}

Next, in preparation for applying the PG test, 
we separate the data into appearance and disappearance datasets and fit independently.  The appearance case effectively has 2 degrees of freedom, while the disappearance case has 3.  Figure ~\ref{fig:disglobal} and Fig. ~\ref{fig:appglobal} show a consistent pattern of Wilks's theorem not being a conservative estimate of $p$-value for these models.  In both cases, the deviation from Wilks's theorem increases with an increased number of bins.

Figure ~\ref{fig:appglobal} also shows a zero bin that is not continuously connected to the rest of the distribution.  This is due to the fact that, in some cases, the best fit will be a negative amplitude.  As this is a $\sin^2(mx)$ fit with a strictly non-negative amplitude, if the ``true'' best fit is a negative amplitude, the best fit in our range will be an amplitude of zero, which is precisely the null (no new physics) case.  This happens in a small number of cases (on the order of 10\%) depending on the fit parameters).  This effect is masked in Fig.~\ref{fig:disglobal}, as it would only show up if both disappearance channels were best fit to null.  It is not shown here, but when disappearance channels are examined independently, they show a similar effect.  This has implications for the PG-test, discussed below.

\begin{figure}
    \centering
    \includegraphics[width=\linewidth]{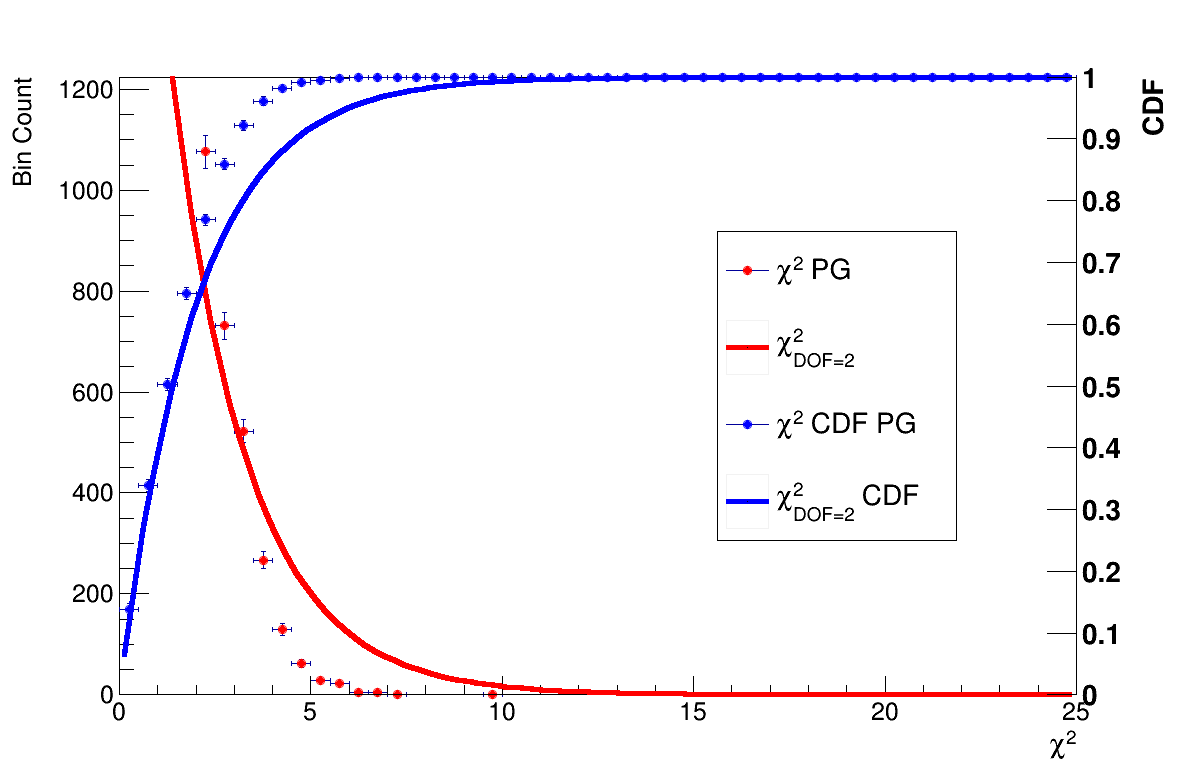}
    \includegraphics[width=\linewidth]{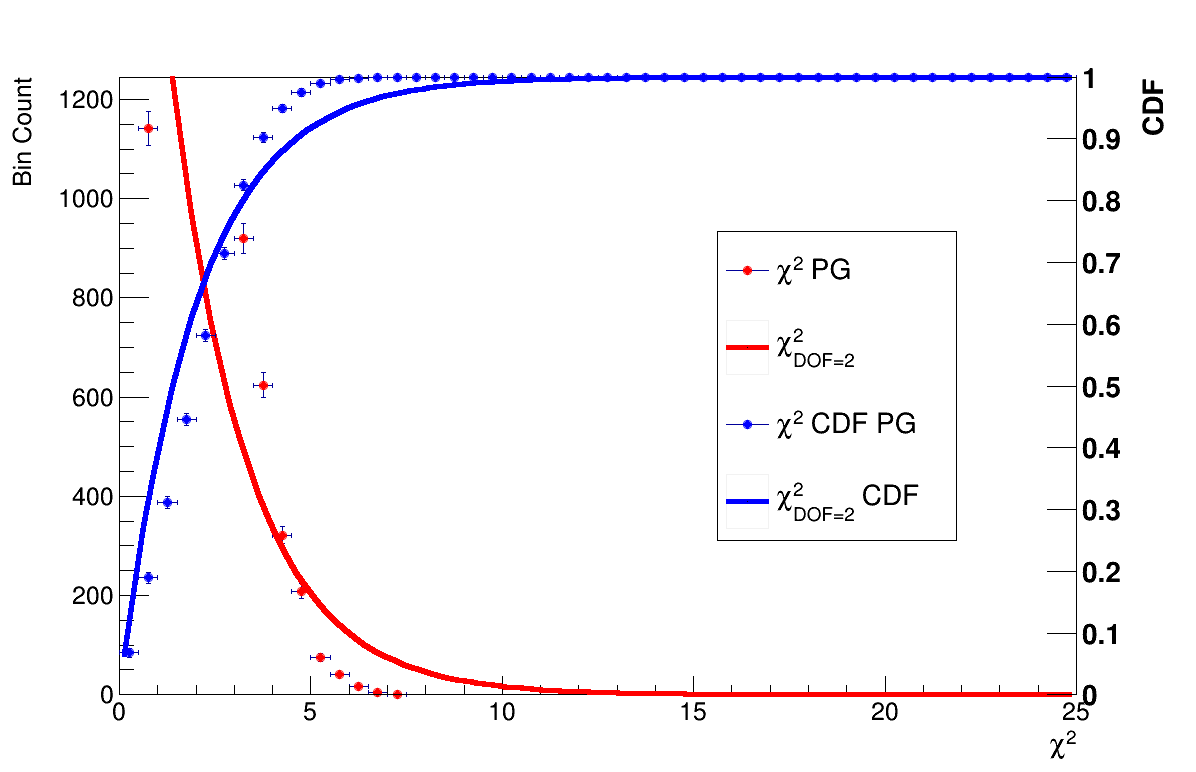}
    \caption{``PG'' fits for the oscillatory model.  Like the other oscillatory cases, Wilks's theorem does not give a good prediction of the test statistic.\label{fig:pgglobal}}
\end{figure}

\subsubsection{More or Less Conservative?  \label{sec:bothdir}}
It is possible for the PG test to be either more or less conservative than Wilks's theorem would suggest based on the underlying model.  
Naively, one might expect the tension to come from one source of $m_{dis}$ being different from $m_{app}$ and the $\chi^2$ wells being abnormally deep due to the effect of checking many independent frequencies and taking the best, described above.  One can visualize this by looking at Fig.~\ref{fig:oscsample} and imagining a best fit point with 2 or 3 minima.  If one type of experiment prefers 2 minima, and one prefers 3, there will be tension.
In other cases, due to the fact that the model has strictly positive amplitudes, the data can under-fluctuate and result in all non-zero amplitudes creating a worse fit.  When this happens, the best fit point is exactly at the null hypothesis, and the best possible $\Delta\chi^2$ is exactly 0.  
Depending on model parameters, this seems to occur in 5-20\% of cases.  
This effect is visible in the jump in the 0 bin for the appearance parameters, but it has an additional effect on the PG test.  
When one of these under-fluctuations happens, it exerts a leftward effect on the PG-test.  
When separated into appearance and disappearance, the amplitudes for all experiments float independently, while when fit together, it is possible to set the appearance and one of the disappearance sets to at or near 0 while allowing the other to float.  
This means that the gain in $\Delta \chi^2$ from separating them, assuming the appearance is at or near under-fluctuated, will just be how much better a single amplitude at the best fit $m_{dis}$, which is a value much closer to 0 than the 2 degrees of freedom expected by the PG test.  This behavior results in multiple intersections of the realizations with the expected Wilks distribution, as visible in Fig.~\ref{fig:pgglobal}.

\subsubsection{Allowed Regions}

Most analyses have the ultimate goal of both discovering something new (``excluding the null'') and characterizing what that new thing might be (``setting allowed regions'' of parameters).  One way to visualize both of these procedures is with an allowed region plot.  Such a plot is constructed by examining a set of data and finding the best fit model parameters $\vec{\theta}_{best}$.  Then, we look at the $\chi^2$ distribution and choose a confidence level for our region.  This is our critical value $\chi^2_{crit}$.\footnote{There are some major subtleties in interpretation here with respect to Bayesian and Frequentist statistics.  We will not discuss them, as this is a practical procedure for setting and interpreting confidence regions}  If Wilks's theorem holds, these values can be read directly off of Fig.~\ref{fig:chi2cdf}.  Then, all models which satisfy:
\begin{equation}
    \chi^2(\vec{\theta}) - \chi^2(\vec{\theta}_{best}) < \chi^2_{crit}
    \label{eq:critdef}
\end{equation}
are said to be in the allowed region for that confidence level.  If that region includes the null ($\vec{\theta}_0$), then we know that we have not excluded the null to that confidence.  Note that Eq.~\ref{eq:critdef} is our familiar $\Delta \chi^2(\vec{\theta}_{best})$ from Eq~\ref{eq:delchi2def} if $\vec{\theta}=\vec{\theta}_0$

In this, final, result section, we will look at some of these allowed regions.  They represent three ways of looking at the global fit of the oscillatory toy model.  The horizontal axes are the appearance ($4A_{e}A_{\mu}$) and disappearance ($4(1-A_{e})A_{e}$, $4(1-A_{\mu})A_{\mu}$) amplitudes that are explored.
These are convenient ways to express the amplitude of appearance and disappearance.  This means that the left-hand side of the plot corresponds to the case where the amplitudes are zero so there is no oscillation.   The vertical axes represent $m$, the frequencies explored.  These bounds change the statistical behavior of the model as they may change the allowed frequency.  Wilks's theorem expects unbounded model parameters, and so, by adding bounds, the degree of violation is changed.  If $m=0$ then there will be no oscillation.  

\begin{figure}
    \centering
    \includegraphics[width=\linewidth]{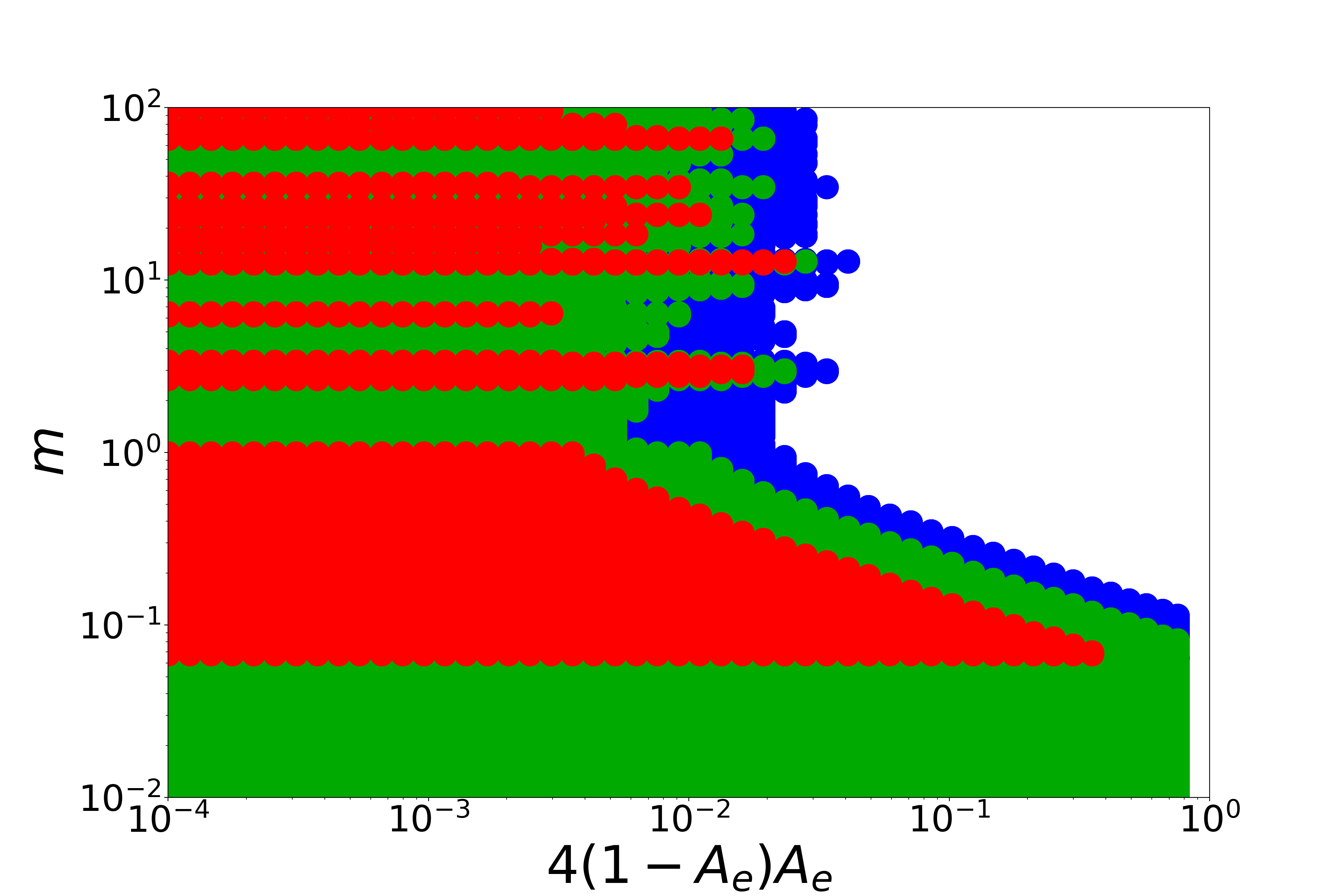}
    \caption{ A single trial's allowed region  The electron disappearance region is shown. \label{fig:toyexclusionsingle}}
\end{figure}

Figure~\ref{fig:toyexclusionsingle} shows the allowed regions for 1 of the 10,000 trials used to produce the $\Delta \chi^2$ histograms above.  The points involved in the scan over parameter space are visible.  The red/green/blue regions correspond to the 90/95/99\% confidence regions based on Wilks's theorem assuming 2 degrees of freedom.  All three of these regions extend to the lefthand side of this plot, which means they include the null.  Therefore, we can conclude that Wilks's theorem will assign a $p$-value greater than 0.1 to this best fit point.  As there was no oscillation in this model, it is good that we did not find any.

\begin{figure}
    \centering
    \includegraphics[width=\linewidth]{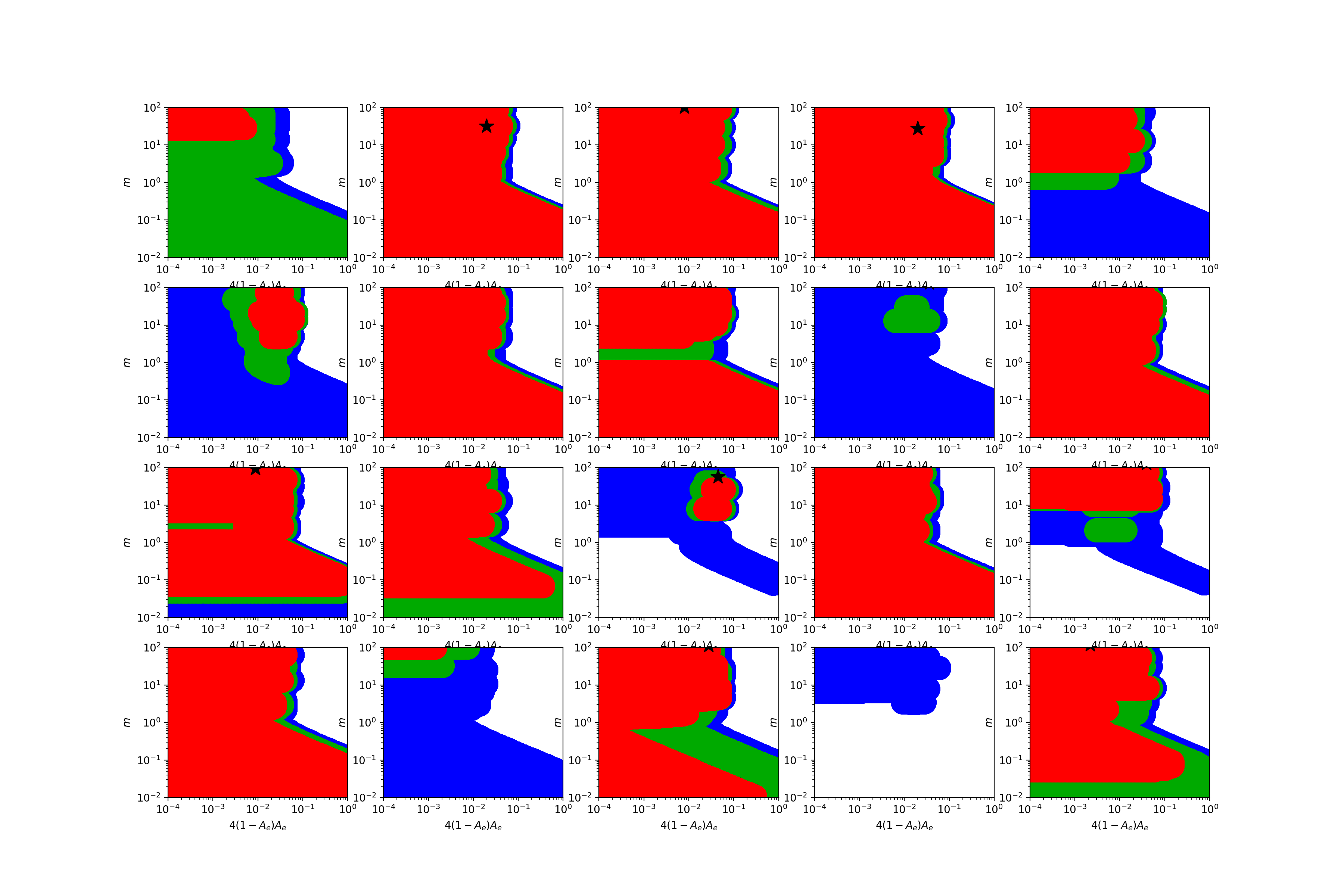}
    
    \caption{ The allowed regions for 20 randomly selected realizations of the toy model.  The red/green/blue regions correspond to the 90/95/99\% confidence regions based on Wilks's theorem assuming 2 degrees of freedom.  The data are shown as electron disappearance (top) muon disappearance (middle), and electron appearance (bottom). \label{fig:toyexclusions}}
\end{figure}

\begin{figure}[tb]
    \centering
    \includegraphics[width=\linewidth]{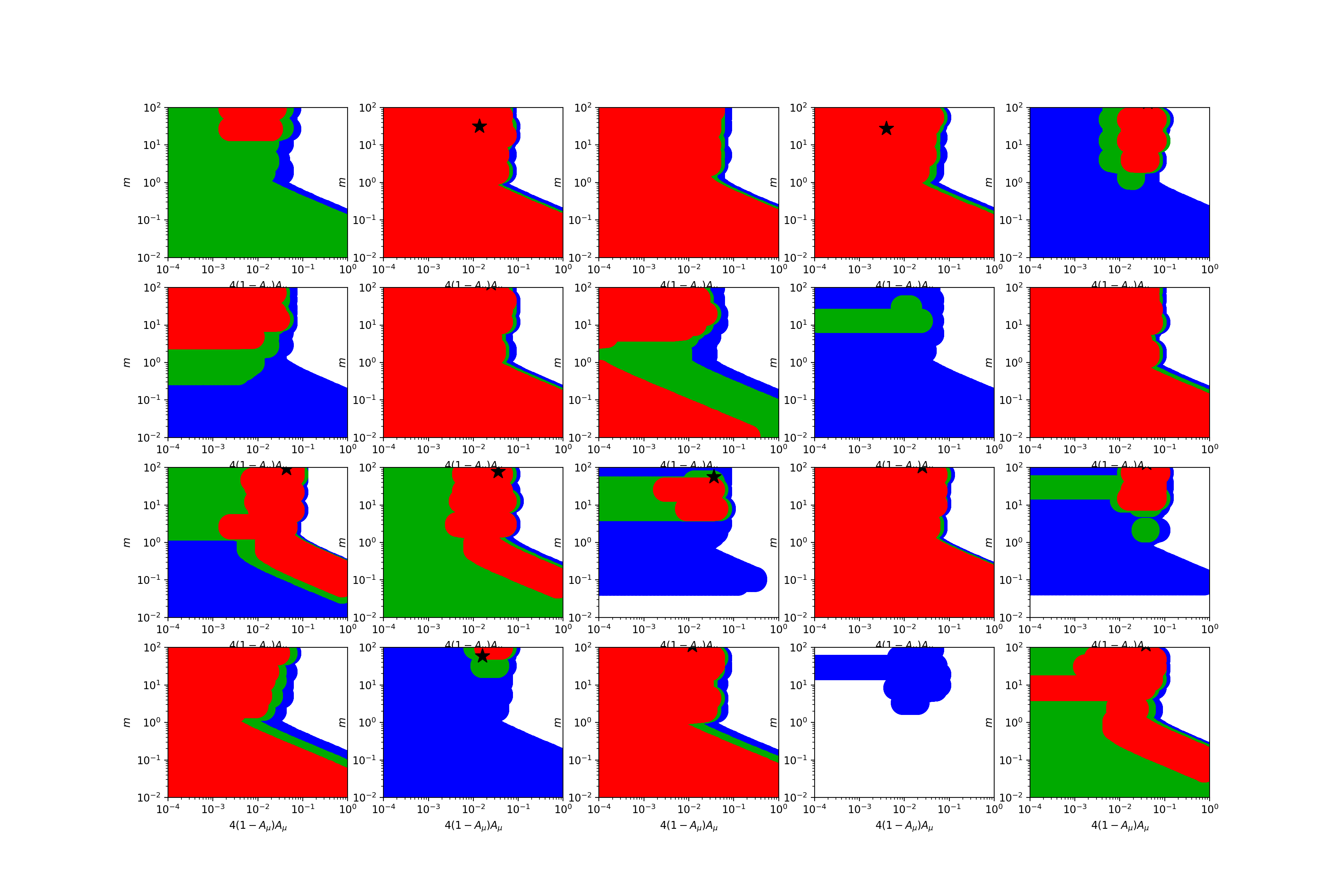}
    \includegraphics[width=\linewidth]{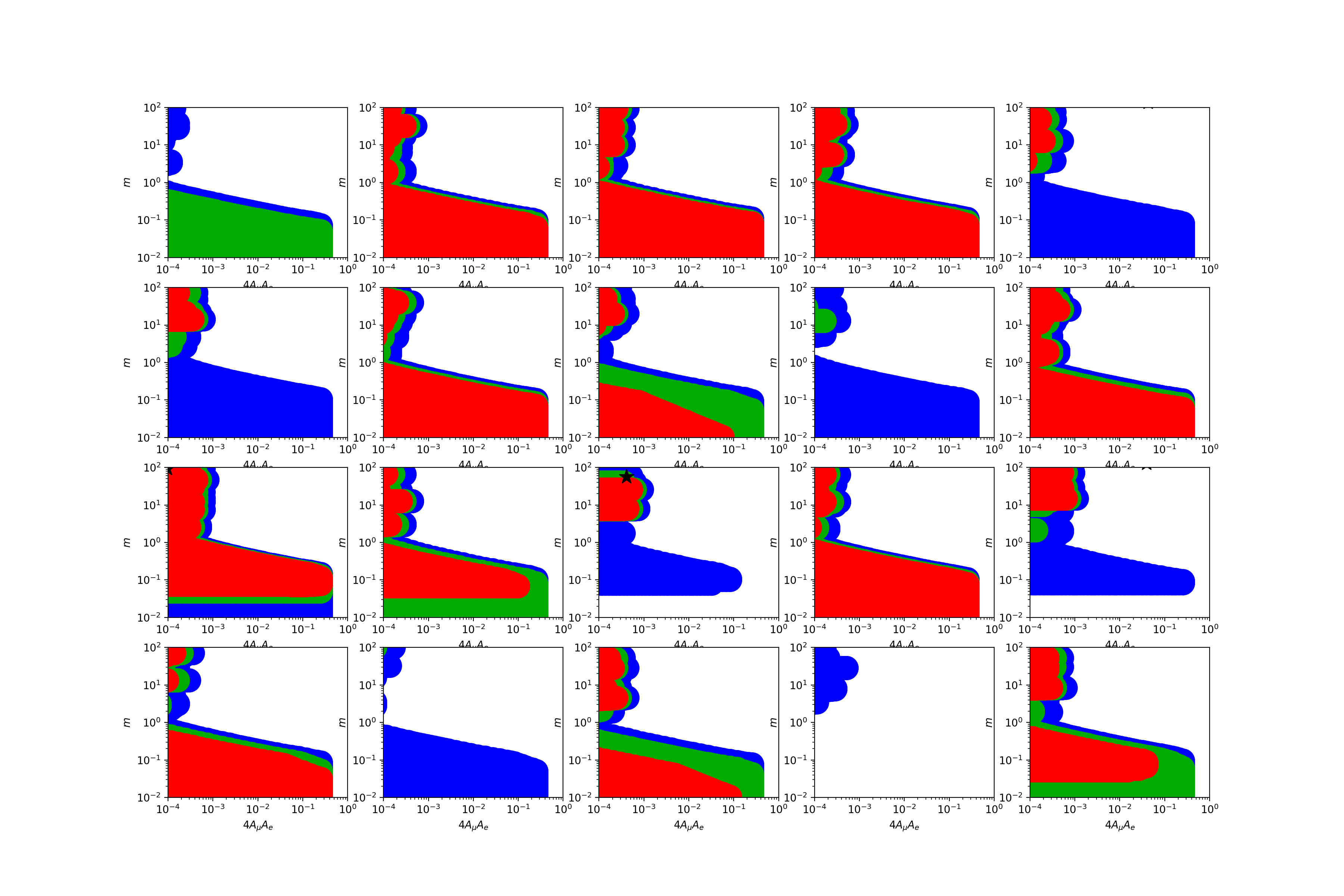}
    \caption{ The allowed regions for 20 randomly selected realizations of the toy model.  The red/green/blue regions correspond to the 90/95/99\% confidence regions based on Wilks's theorem assuming 2 degrees of freedom.  The data shown are muon disappearance (top), and electron appearance (bottom). \label{fig:toyexclusionsother}}
\end{figure}

Figure~\ref{fig:toyexclusions} shows these regions for electron disappearance in 20 different random trials.  Figure~\ref{fig:toyexclusionsother} shows the same data in the muon disappearance and electron appearance parameters.  One trial seemed to produce the correct answer, but if we look at 20 at a time, we can get a better sense of how our statistical procedure is performing.  When the models are obeying Wilks's theorem, we expect them to exclude the the left hand side of the plot 10/5/1\% of the time.  Looking at these 20 realizations we might suggest that the null is excluded more often than expected.  The red does not touch the left hand side of the plot in 4 of the 20 random cases, when we would expect 2.  Looking at a full test such as Fig.~\ref{fig:delglobal} more definitively proves the inapplicability of Wilks's theorem, but this provides a visualization of the problems of a test statistic whose true distribution lies to the right of the ideal $\chi^2$. 

\subsubsection{Exclusion and Allowed Regions}

\begin{figure}[tb]
    \centering
    \includegraphics[width=\linewidth]{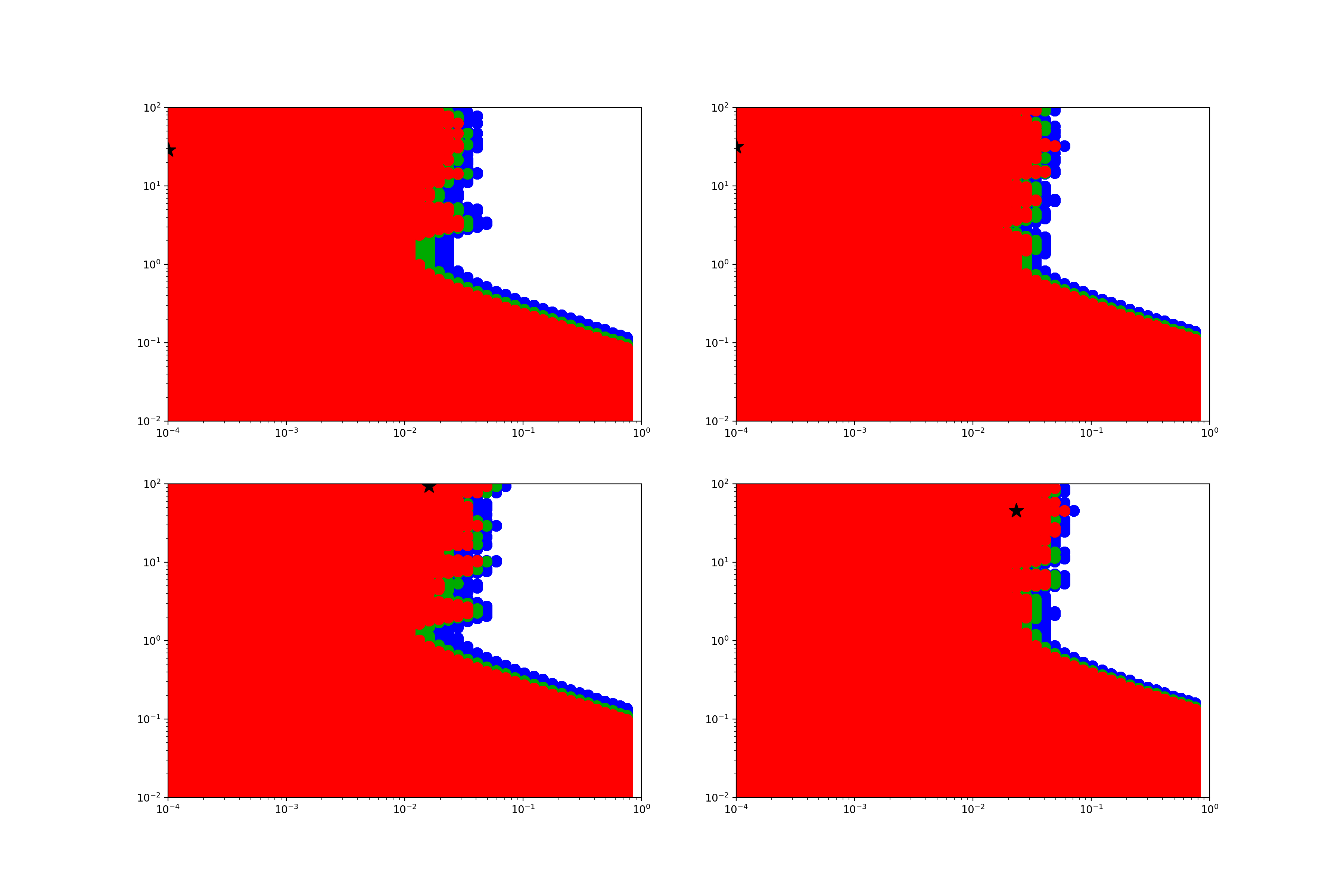}
    \includegraphics[width=\linewidth]{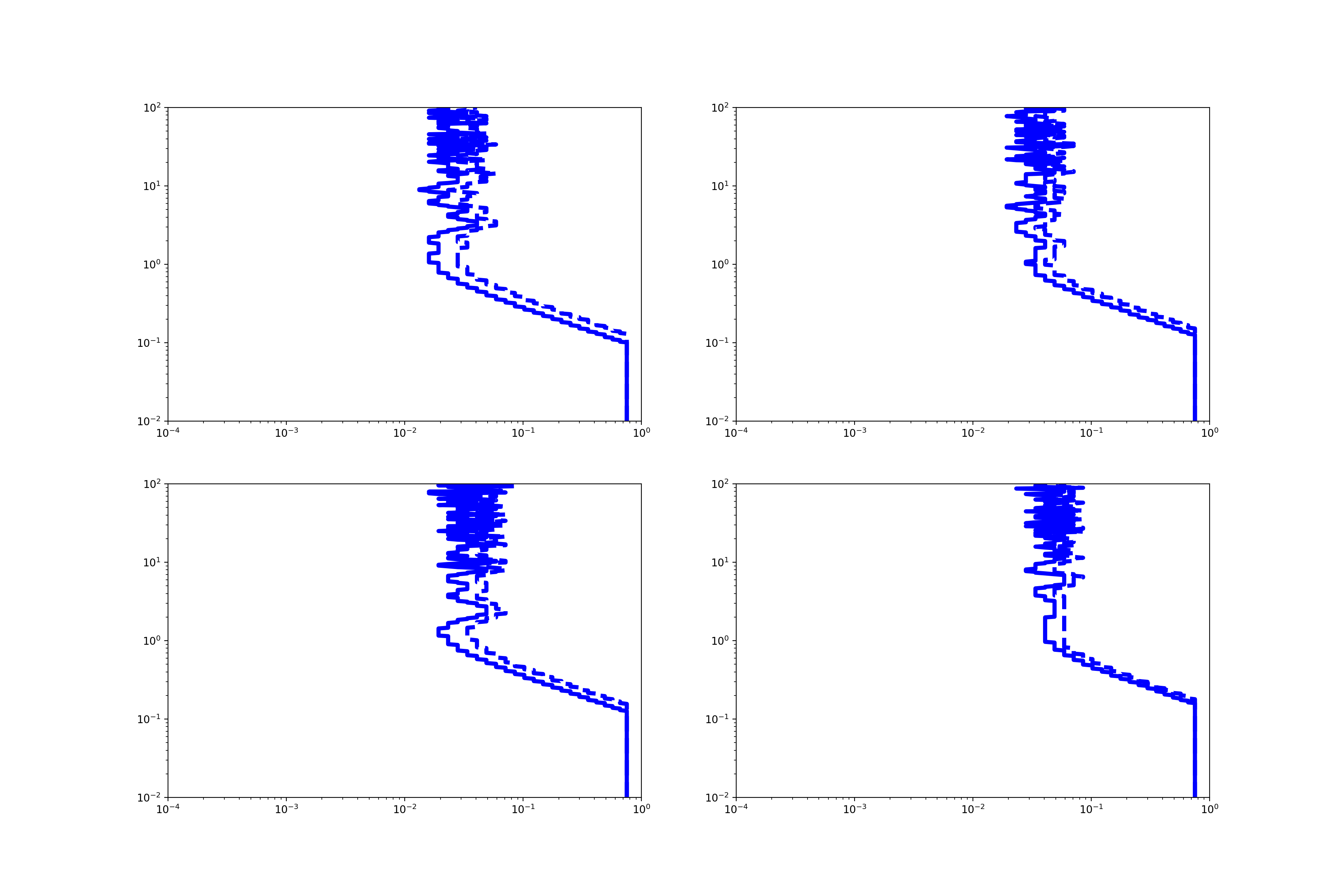}
    \caption{ The electron disappearance allowed plots of the first 4 toy experiments (Top).  The 90\% exclusion curve (Bottom).  The Solid Exclusion curve assumes Wilks's theorem, while the dotted curve uses a $\chi^2$ threshold of 9.5 based on the 90\% level of the CDF in Fig.~\ref{fig:delglobal}. \label{fig:toycurves}}
\end{figure}

When analyses do not discover something new, they will typically report an exclusion curve as in Fig.~\ref{fig:toycurves}.  Such a plot is constructed by examining a set of data and comparing the $\chi^2$ value at each set of model parameters to $\chi^2_{0}$.  The curve is drawn by selecting a value of $\chi^2$ corresponding to the probability of a value that large or larger.  This is our critical value $\chi^2_{crit}$.  If Wilks's theorem holds, these values can be read directly off of Fig.~\ref{fig:chi2cdf}.  Then, all models which satisfy:
\begin{equation}
    \chi^2(\vec{\theta}) - \chi^2_{0} > \chi^2_{crit}
    \label{eq:critdefallowed}
\end{equation}
are said to be in the excluded region for that confidence level.  Typically, for models close to the null, the left-hand side of Eq.~\ref{eq:critdefallowed} varies monotonically with the amplitude of the new physics, and all models above a given amplitude are excluded.

This procedure is strongly connected to the allowed region procedure.  If the best fit is at or near the null, then the exclusion curve will be near the corresponding allowed region curve.  This can be seen by comparing the solid line in the bottom panel of Fig.~\ref{fig:toycurves} to the red regions in the top panel.  

But, as we discussed, Wilks's theorem does not apply, and so 
$\chi^2_{crit}$ must be set by the CDF generated from trials.  Figure~\ref{fig:delglobal} allows us to read this value as approximately 9.5 (Where the CDF of the top panel, which is the 20 bins case as shown here, crosses 90\%).  Using this as our critical value produced the dotted exclusion line in Fig.~\ref{fig:toycurves}, which are to the right of the solid lines (note the logarithmic scale).  As these curves exclude less space to their right, they are less sensitive, which means that Wilks's theorem would have caused us to over-report the sensitivity of our experiment, which is consistent with the allowed regions under including the null in the previous section\footnote{Properly identifying allowed regions requires much more computation time and many fits at each set of model parameters.  See Ref~\cite{Feldman_1998} for a full method of producing allowed regions.}

\subsection{Conclusions}

Wilks's theorem is a useful tool but can be inapplicable to many types of models.  Specifically, it does not accurately describe the statistics of the sterile neutrino global fits.  Both the allowed regions and the tension are affected by these deviations, and both can have their $p$-values underestimated by Wilks's theorem.  When interpreting physics analyses, therefore, it is important to rigorously check the applicability of any simplifying assumptions such as Wilks's theorem.  Pseudo experiments, as shown in this paper, are an effective, if computationally intensive, way to test and correct these simplifying assumptions.  This is especially important when the assumptions will result in an analysis incorrectly assigning a high significance to a set of data, but error on the significance in either direction negatively impacts the quality of information from the analysis.

\subsection{Further Reading}

It is beyond the scope of this paper to discuss the various solutions to this problem, but the interested reader may wish to look into the following references: Ref.~\cite{Diggle1984} for a rigorous treatment of significance, Ref.~\cite{Feldman_1998} for a unified treatment of parameter estimation and significance testing, and Refs.~\cite{Berger1994},~\cite{Cousins:1991qz}, ~\cite{conrad2002coverage},~\cite{Tegenfeldt_2005},~\cite{Baxter_2021}, and~\cite{acero2022profiled} for treatment with more realistic experimental considerations (especially systematic uncertainties).

\section{Acknowledgements}
J. Hardin would like to thank J. Conrad, S. Robinson, and M. Shaevitz for comments and edits.  JMH is supported by NSF grant PHY-1912764.

\bibliographystyle{apsrev}
\bibliography{toymodels}

\end{document}